\documentclass[aps,pre,twocolumn,groupedaddress,showpacs]{revtex4}
\usepackage{latexsym}
\usepackage{amssymb}
\usepackage{graphicx}
\newcommand{\W}{8cm}

\begin{document}


\title{Pore scale mixing and macroscopic solute dispersion regimes in polymer flows inside 2D model
networks.}



\author{Maria Veronica D'Angelo$^{1,2}$}\email{vdangelo@fi.uba.ar}
\author{Harold Auradou$^1$}\email{auradou@fast.u-psud.fr}
\author{Catherine Allain$^1$}
\author{Jean-Pierre Hulin$^1$}\email{hulin@fast.u-psud.fr}

\affiliation{$^1$Laboratoire Fluides, Automatique et Syst{\`e}mes 
Thermiques, UMR 7608, Universit{\'e}s Pierre et Marie Curie-Paris 6 et Paris-Sud, 
B{\^a}timent 502, Campus Paris Sud, 91405 Orsay 
Cedex, France\\$^2$Grupo de Medios
Porosos, Facultad de Ingenieria, Paseo Colon 850, 1063, Buenos
Aires, Argentina}



\date{\today}

\begin{abstract}
A change of solute dispersion regime with the flow velocity has been studied both at the macroscopic and pore scales in a transparent array of capillary channels using an optical technique allowing for simultaneous
 local and global concentration mappings. Two solutions of different polymer concentrations ($500$ and $1000$~ppm) have been used at different P\'eclet numbers. 
At the macroscopic
scale,  the displacement front displays a diffusive spreading: for  $Pe\  \leq \ 10$, the  dispersivity $l_d$
 is constant with $Pe$ and increases with the polymer concentration;  for $Pe \ > \ 10$, $l_d$
  increases as $Pe^{1.35}$ and is similar for the two concentrations. At the local scale, 
  a time lag between the saturations of channels parallel and perpendicular to the mean flow has
  been  observed and studied as a function of the flow rate. These local measurements suggest that the change of dispersion regime is related to variations of the degree of mixing at the junctions. 
For  $Pe \ \leq \ 10$, complete mixing   leads to pure geometrical 
 dispersion  enhanced for shear thinning fluids; for $Pe >10$ weaker mixing results 
 in higher correlation lengths along flow paths parallel to the mean flow and in a combination 
 of geometrical and Taylor  dispersion. 
\end{abstract}


\maketitle

\section{Introduction}
\label{intro}
The problem of solute transport in porous media is relevant to
 many  environmental, water supply and industrial
processes~\cite{bear72,dullien91}. In addition, tracer dispersion is a useful tool to analyse  porous
media heterogeneities~\cite{charlaix88}.
Solute dispersion is often measured  by monitoring solute concentration
variations at the outlet of the sample following a pulse or step-like  injection at the inlet: 
Understanding 
fully the dispersion mechanisms requires however informations on the local concentration
 distribution and on  local mixing at the pore scale. These goals can be reached by using 
 NMR-Imaging, CAT-Scan or acoustical techniques~\cite{wong99} : 
however, such techniques are either costly or have a limited resolution. In addition, they often 
put strong constraints on the  characteristics of the  fluid pairs. We have chosen instead in the present work to use a $2D$  transparent square network of ducts of  random widths allowing  for easy visualizations of mixing and transport processes at the local scale.  Similar systems were previously
 used successfully  to investigate two phase flows~\cite{zarcone83} and miscible displacements of Newtonian fluids~\cite{Birovljev94}.\\
The key feature of the present experiments is to combine  macroscopic
and local scale measurements to estimate the influence of pore scale processes
on dispersion.
  For this purpose, dye is used as a 
solute and high resolution maps of the relative concentration distribution are obtained
through calibrated light absorption measurements. This allows one to determine quantitatively
the global dispersion  coefficient $D$ and the dispersivity $l_d$ 
 while measuring simultaneously  the time lag between the invasion of channels parallel and 
transverse to the mean flow.  Further informations are also obtained 
from the complex geometry of an isoconcentration line. \\
This approach have allowed us to observe  a change of dispersion regime for a P\'eclet
number of the order of $10$ while the local measurements suggest interpretations in terms of  variations of the degree of mixing at the junctions. This  confirms previous suggestions~\cite{grubert01,park01} on the influence of mixing at the pore scale on macroscopic dispersion.\\
Another feature of our experiments is the use of shear thinning polymer solutions of the type encountered in  many  industrial processes in petroleum, chemical and civil
engineering~\cite{sorbie89}. In addition to these practical applications,
previous measurements at the laboratory scale on glass bead packings have shown~\cite{paterson96,freytes01} a significant enhancement of tracer dispersion compared to the 
Newtonian case. This enhancement depends on the polymer concentration 
and will represent  a useful   additional input for our interpretations. 
\section{Dispersion mechanisms in 3D porous media and 2D networks}
\label{dismech}
In homogeneous $3D$ porous media, the macroscopic
 concentration $\overline C$ of a tracer  (i.e. averaged over a representative 
 elementary volume) satisfies the convection-diffusion equation :
\begin{equation}\label{eq:condiff}
\frac{\partial {\overline C(x,t)}}{\partial t} = U\frac{\partial
{\overline C(x,t)}}{\partial x} +D \frac{\partial^2{\overline
C(x,t)}}{{\partial x}^2}
\end{equation}
where $D$ is the longitudinal dispersion coefficient,  $U$ the 
mean velocity of the fluid (parallel to $x$) and $\overline C$ is 
assumed to be constant in a section of the sample normal to  {\bf U}. 
In the following, equation~(\ref{eq:condiff}) is shown to be also valid
 for the transparent model used in the present work.\\
The value of $D$  is determined by two main physical mechanisms : molecular diffusion 
and advection by the complex velocity field inside the medium 
(the local flow velocity varies both inside individual flow channels and
 from one channel to another).  The relative order of magnitude of these two effects is 
 characterized by  the  P{\'e}clet number: $Pe = {Ua}/{2\ D_m}$ ($D_m$ is the molecular diffusion 
 coefficient and  $a$ a characteristic length of the medium, here the average 
 channel width).\\
Various dispersion regimes are observed in usual porous media~\cite{bear72}. At very low 
P{\'e}clet  numbers ($Pe < 1$), molecular diffusion  is dominant and smoothes out  local concentration variations.\\
At higher $Pe$ values, the distribution of the channel widths induces short range 
variations  of the magnitude and the direction of the local velocity. 
One can then consider that tracer particles experience a random 
walk inside the pore volume with a velocity varying  both in magnitude and direction
relative to the mean flow velocity ${\bf U}$.  The typical length $l$ of the channels  represents  the length of the steps and their characteristic duration $\tau$ is $\tau \approx l/U$.  
A classical feature of random walks is that the corresponding diffusion coefficient (here equal to $D$) satisfies $D \propto  l^2/\tau = U l$. The 
 proportionality constant depends both on the disorder of the medium and on the rheology of the fluid.  In this so-called {\it geometrical dispersion}, the coefficient $D$ should then be proportional to $U$. \\
The $2D$ networks of interest in the present work have  several specific features and 
$D$ should be  very much  influenced  by the redistribution of the incoming tracer between the  channels  leaving a junction. This redistribution strongly depends on the 
 local   structure of the flow field and on the P{\'e}clet number~\cite{mourzenko02,berkowitz94}. 
At low $Pe$ values, the transit time through a junction  is large enough for tracer to cross 
streamlines by molecular  diffusion and one may assume a perfect mixing. 
In the other limit $Pe>>1$, molecular diffusion is negligible: the path of the tracer particles
coincides with the flow lines and is determined by the flow  field in the junctions and by the location of the particles in the flow section. \\
In the case of $2D$ networks with small variations of the channel apertures, the flow field
is close to that in a periodic square channel with a mean flow parallel to one of the axis: 
the major part of the flow is localized in longitudinal channels parallel to the axis where the 
velocity is high while that in transverse channels is small.  
Tracer particles remain then  inside sequences of channels parallel to the mean flow for a long distance without moving sideways.  Taylor-like dispersion~\cite{taylor53,aris56} similar 
to that encountered in capillary tubes, between parallel plates or inside periodic structures may then develop: 
 it corresponds to  a  balance between  (a) spreading due to velocity gradients between the 
 centers of the flow channels and their walls  and (b) molecular diffusion across the flow lines. 
The increased dispersion in 2D periodic networks  when flow is parallel to one of the axis~\cite{grubert01} may for instance result from a combination of geometrical and Taylor dispersion.\\ 
Another issue of the present work is the influence of the shear-thinning properties
of the fluids.  They may  influence the dispersion process in different ways:  on the one hand,
when the viscosity $\mu$ decreases with the shear rate $\dot{\gamma}$, 
 the velocity profile in individual channels becomes flat in the center of the channel. 
  In simple geometries like capillary tubes,  the corresponding Taylor dispersion coefficient $D$ is then lower than for a Newtonian fluid although one has still $D \propto Pe^2$~\cite{vartuli95}. 
 On the other hand, numerical investigations suggest that the flow of shear thinning fluids is localized in a smaller  number of flow paths than for Newtonian fluids~\cite{shah95,fadili02}. 
 As a result, geometrical dispersion reflecting the distribution of the local velocities should  increase   as is indeed observed experimentally~\cite{paterson96,freytes01}. 
A key point is here the influence  of the fluid rheology on tracer mixing at the intersections
 between channels: to our knowledge, no previous experimental or numerical work has dealt 
 with this issue which is an important point. 
\section{Experimental set-up and procedure} \label{sec:exp}
\subsection{Experimental models and injection set-up} \label{subsec:model}
The model porous medium is  a  two dimensional square
network  of channels of random aperture: these models have been
 realized by casting a transparent resin on a
photographically etched mold as  described in
ref.~\cite{zarcone83}. The model contains a square network of $140 \times140$
channels with an individual length equal to $l \ = \ 0.67$~mm and a depth
of $0.5$~mm; the width follows  a discrete, log-normal
distribution with $7$ values between $0.1$ and $0.6$~mm (the average width is $\bar{a}=0.33$ 
and its standard deviation  $\sigma(a)\ =\ 0.11$~mm). The mesh size of the network is
equal to $1$~mm.  The overall size of the model  is
$150 \times 140$~mm and the two facing lateral sides are sealed. On the two others, the channels are directly connected to
the outside. The total pore volume is close to $6.09\times 10^3$~mm$^3$.
Following the definition of Bruderer {\it et al}~\cite{bruderer01}, the degree of heterogeneity of the 
network can be caracterized by the normalized standard deviation $\sigma(a)/\bar{a}$. 
In the present work : $\sigma(a)/\bar{a} \simeq 0.33$.
\begin{figure}[h!] 
\noindent\includegraphics[width=\W]{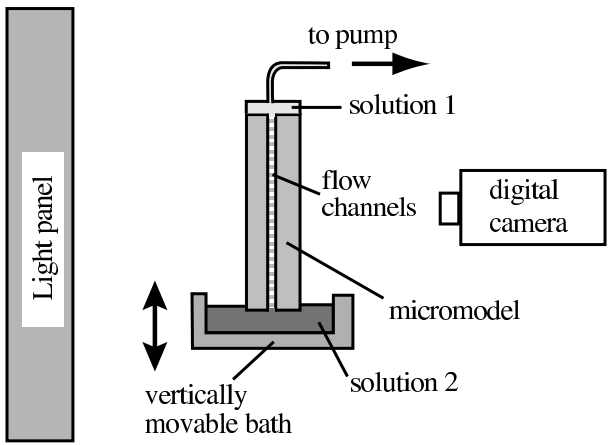}
\caption{Experimental setup for miscible displacement
measurements in 2D micromodels.}
\label{fig:fig1}
\end{figure}
\subsection{Fluid preparation and characterization}
\label{rheo}
The fluids used in the experiments are shear thinning water-polymer solutions.
The shear thinning fluids are solutions of $500$ to $1000$~ppm of high 
molecular weight scleroglucan in high purity water
(Millipore - Milli-Q grade).  Scleroglucan (Sanofi Bioindustries,
France)  is a polysaccharide  with a semi rigid molecule
(persistence length  $\simeq180\times 10^{-9}$~m);  
it has been selected because it
is electrically neutral and its characteristics are therefore  independent
of the ion (and dye) concentration. All solutions are protected
from bacterial contamination by adding
$0.2$~g/l of $NaN_3$.
In all experiments, the injected and displaced fluids are
identical except for a small amount of Water Blue dye which is added 
to one of the solutions at a concentration
of $200$~ppm by weight to allow for optical concentration measurements.\\
The molecular diffusion coefficient of the dye is determined independently in
pure water by means of Taylor dispersion measurements performed separately in a 
capillary tube (the measured 
value of $D_m$ was close to $6.5 \times 10^{-4}$~mm$^2$.s$^{-1}$.)\\
The rheological properties of the scleroglucan solutions have been characterized using
 a {\it Contraves LS30} Couette rheometer with shear rates $\dot{\gamma}$ ranging from
  $0.016$~s$^{-1}$ up to $87$~s$^{-1}$. The rheological properties of the solutions have been verified
to be constant with time within experimental error (over a time lapse of $3$ days) and 
 identical for dyed and transparent
solutions with the same polymer concentrations. The variation of the effective viscosity $\eta$
 with $\dot{\gamma}$ was then adjusted by a 
Carreau function : 
\begin{equation}\label{powervisc}
\eta = \frac{1}{(1+ (\frac{\dot{\gamma}}{\dot{\gamma_0}})^2)^{\frac{1-n}{2}}} (\eta_0 - \eta_{\infty}) + \eta_{\infty}.
\end{equation}
The values of these different rheological parameters for the polymer solutions used in the
present work are listed in Table~\ref{tab:tab1}; determining $\eta_{\infty}$ would have 
required values of $\dot{\gamma}$ outside the measurement range so that we assume
 that  $\eta_{\infty}\,=\,10^{-3}$~Pa.s (the value for the solvent i.e. water). 
 In Eq.(\ref{powervisc}), $\dot{\gamma_0}$ corresponds to a crossover between two regimes. 
 On the one hand, $\dot{\gamma}<\dot{\gamma_0}$, the viscosity $\eta$ tends towards $\eta_0$, 
 and the fluid displays a Newtonian behavior ("Newtonian plateau" regime). On the other hand, 
 for $\dot{\gamma}\gg\dot{\gamma_0}$, the effective viscosity decreases with the shear rate 
 following the power law  $\eta \propto {\dot{\gamma}}^{(n-1)}$ ($n = 1$ for a Newtonian fluid.)
 It should finally be noted that 
the high effective viscosity  of these solutions at low shear rates avoids the appearance of
buoyancy induced instabilities at low shear rates and helps stabilize the fluid displacement.
\begin{table}[htbp]
\begin{tabular}{lccc}
Polymer Conc. & $n$ & $\dot{\gamma_0}$ & ${\eta}_0$\\
 ppm          &     & $s^{-1}$       & $mPa.s$\\
\\
$500$ & $0.38 \pm 0.04$ & $0.077 \pm 0.018$ & $410 \pm 33$ \\
$1000$ & $0.26 \pm 0.02$ & $0.026 \pm 0.004 $ & $ 4500 \pm 340$ \\
\end{tabular}
\caption{Rheological parameters of scleroglucan solutions used
in the flow experiments.}
\label{tab:tab1}
\end{table}
\subsection{Fluid injection and flow visualization}
The model is placed vertically with its open sides horizontal :
the upper  side is fitted with a leak tight adapter allowing one to
suck  fluid upwards. The lower open
side is initially slightly dipped into a bath of one of the liquids 
and is saturated with this fluid by pumping it slowly upwards. After switching off the pump, 
the bath can be lowered until it does not touch any more the model
(Fig.~\ref{fig:fig1}). 
The bath is then completely
emptied, refilled with the other fluid and raised to its initial position.
Finally, the first fluid is sucked upwards at the upper end of the model
 by the syringe pump. This allows one to
obtain a front of the displacing fluid which is initially perfectly
straight.  The lower bath rests
upon computer controlled  electronic scales for monitoring
the amount of fluid which has entered the model. 
The flow rates used in the experiments correspond to mean front
velocities between $0.005$ and $2.5$~mm.s$^{-1}$.\\
The model is illuminated from the back by an electronic light panel and images 
are acquired  by a $12$ bits,
high stability, digital camera with a $1030 \times 1300$ pixels
resolution and then recorded by a computer. The pixel size is
$0.14$~mm or $0.4$ times the mean width of the channels: this
allows one to discriminate between the various regions of the
pore space. Typically $100$ images are recorded for each
experimental run at time intervals  from
$2.5$ to $700\,s$.
\subsection{Image analysis procedure}
The images are then translated into maps of the relative
concentration using the following procedure.
First, a calibration curve is obtained from images of the model
saturated with $7$ solutions of increasing dye concentration
$C$ starting from zero up to the  concentration used in the experiments. 
The logarithm $Ln(I(C)/I_o)$ of the transmissivity is then plotted
as a function of $C$; $I(C$) and $I_o$ are averages of the
light intensity over the model and correspond respectively to
dye concentrations equal to $C$ and $0$. Due to the non linear absorbance
 effect~\cite{detwiler00}, a better fit is obtained with a third
order polynomial variation of $Ln(I(C)/I_o)$ with $C$ than with
 the linear dependence corresponding to Lambert's law.
This calibration is performed every time  the locations of
the light source and of the models are significantly  changed.\\
For all experiments, reference images are recorded both with
the micromodel  initially saturated with the displaced fluid and, at
long times, when it is  fully saturated with the injected one.
After the fluid displacement has been performed, the local
concentrations are determined pixel by pixel for each image by
means of the calibration curve. This operation is performed only on pixels 
belonging to channels; non flowing domains are not considered. 
Finally, maps of the local {\it relative} concentration of the two fluids are obtained by
normalizing the local concentration between its values in the
initial and final images.
\section{Experimental results}
\label{sec:res}
\subsection{Qualitative observations of miscible displacements}
\begin{figure}[h!]
\includegraphics[width=\W]{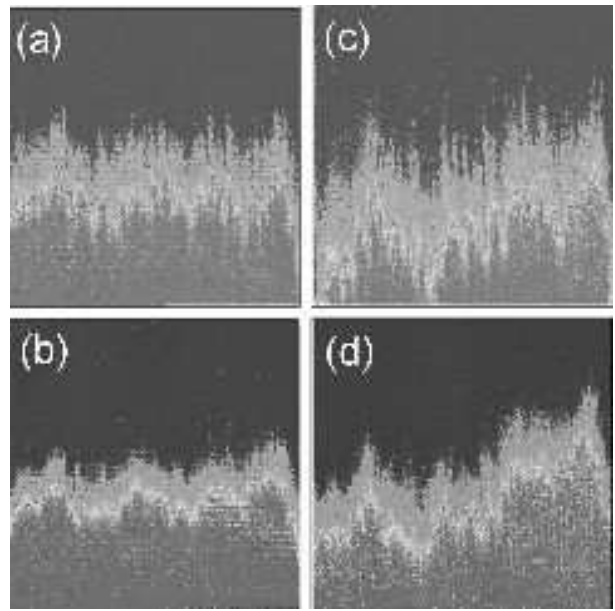}
\caption{Relative concentration maps for experiments using water-scleroglucan
 solutions of respective
concentrations $500$~ppm (a,b) and $1000$~ppm (c,d) for flow
rate values  $Q = 0.075$~ml/mn (b,d) and $Q = 1.5$~ml/mn
(a,c). Grey levels code used in the figure : darkest shade = pure displaced fluid, intermediate
- pure displacing fluid - lightest = intermediate concentrations}
 \label{fig:fig2}
\end{figure}

Figure~\ref{fig:fig2} compares
displacement experiments realized at a same flow velocity for the two water-scleroglucan
solutions.
For a given solution, narrow structures of the front with a lateral
extent of a few channel widths appear when the velocity increases and
reflect local high or low velocity zones.
 For a same flow rate, images obtained with the two solutions
are qualitatively similar:  the size parallel to the flow of
medium scale front  structures (with a width of the order of one
tenth that of the model) is however  larger for the more concentrated
solution  both at low and high velocities.\\
The stability of the displacement with respect to buoyancy driven instabilities has
also been verified by comparing experiments using the same pair of fluids and
exchanging the injected and the displaced fluid~\cite{freytes01}: no quantitative difference
was measured between the two configuraton and no fingering instabilities
appeared in the unstable configuration.\\
While fluid displacement images
provide informations on the  
 front geometry down to a fraction of the 
channel size, we shall see next  that they also allow to determine macroscopic parameters
characterizing the process.
\subsection{Quantitative dispersion measurement procedure}
\label{sec:quant}
Quantitatively, the global  displacement process is
analyzed from  the variations with time and
distance parallel to the flow of a macroscopic concentration
${\overline C}(x,t)$:  ${\overline C}$ is  the average over an interval
$\Delta y$ perpendicular to the flow of the local
relative concentration $C$ for individual pixels located inside
the pore volume. The width  $\Delta y$
is large enough to average out local fluctuations and
small enough avoid the influence of the side walls:   ${\overline C}$ was
found to be independent of $\Delta y$ within experimental error 
when these conditions are verified. Also,  it will be seen below that the results 
are hardly different when   $\Delta y$ corresponds to 
only one channel.\\
\begin{figure}[h!]
\includegraphics[width=\W]{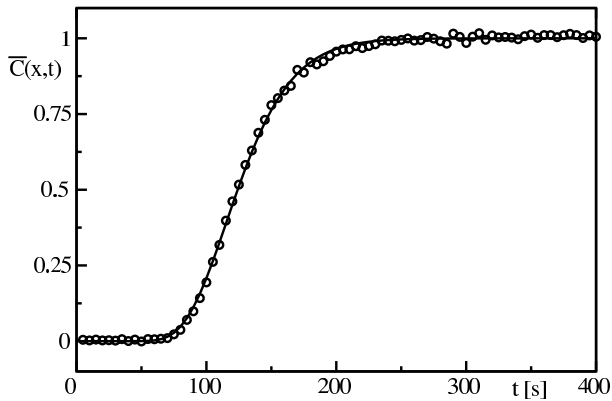}
\caption{Normalized concentration variation 
${\overline C}(x,t)$ as a function of time  for a displacement 
experiment
using a $500$~ppm polymer solution with $Q = 3.75$~ml/mn.
${\overline C}$: averaging interval $\Delta y \simeq 35$~mm located 
 in the central part of the model.
Distance from inlet: $x=80$~mm.
Continuous line : fit by Eq.~(\ref{eq:condiffsol}).}
\label{fig:fig3}
\end{figure}
Figure~\ref{fig:fig3} displays a typical variation of ${\overline
C}$ with time.
For step-like initial concentration variations corresponding to
our experiments, the solution of the convection-dispersion equation~(\ref{eq:condiff}) is:
\begin{equation}\label{eq:condiffsol}
{\overline C} = \frac{1}{2}\left \lbrack  1 - erf \frac{t - 
{\overline t}}{\sqrt{4 \frac{D}{U^2} t}} \right \rbrack
\end{equation}
As seen in Figure~\ref{fig:fig3}, a very good fit of the experimental data
 with this solution (continuous line) is 
obtained by adjusting the two parameters of the equation, namely the mean transit
 time $\overline t$ and the ratio ${D}/{U^2}$ ($= {\overline \Delta t^2}/2\overline{t} $
 where ${\overline \Delta t^2}$ is the centered second moment)\\
\begin{figure}[h!]
\includegraphics[width=\W]{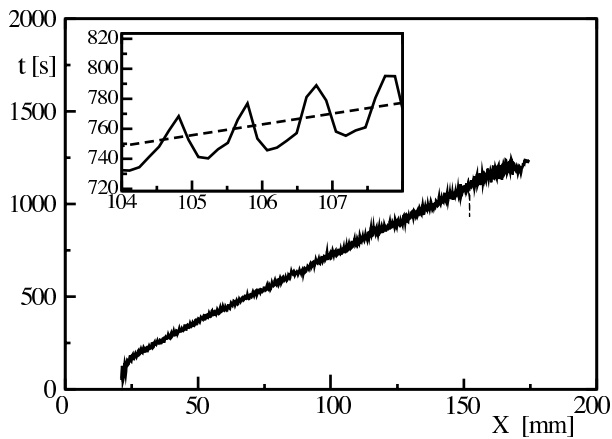}
\caption{Variation of $\overline t$ with the distance $X$ (polymer concentration : $500 ppm$, 
$Q=3.75 ml/min$, $Pe=317$) for dyed fluid displacing clear fluid. Dashed line : clear
 fluid displacing dyed fluid.  Inset :  close-up at the scale of $4$ channels. Dashed line:
  linear regression of the variation of $\overline t$ with $X$.}
\label{fig:fig4}
\end{figure}
Figure~\ref{fig:fig4} displays the variation of the mean transit time $\overline t$ with the distance $X$ 
from the inlet: 
$\overline t$ increases overall  linearly  with distance, indicating that the mean velocity $U$ is constant 
and can be determined by a linear regression of the data.
The inset (magnified view of the curve) shows that $\overline t(X)$ oscillates about the mean trend 
(dotted line) corresponding to the linear regression. These periodic variations are
directly related to the structure of the network and will be discussed in section~\ref{secI}.
\begin{figure}[h!]
\includegraphics[width=\W]{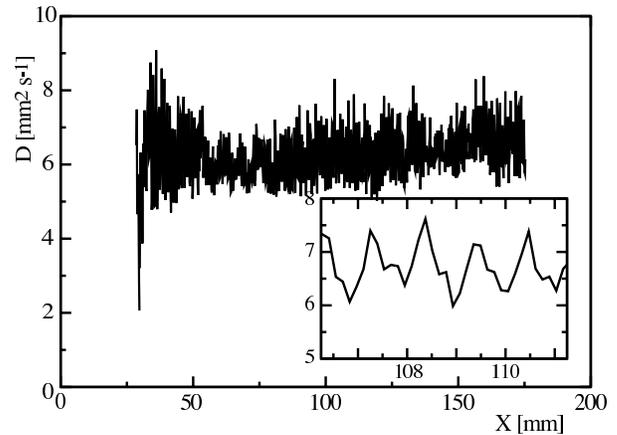}
\caption{Variation of $D$ with the  distance $X$ for dyed fluid displacing clear fluid (Polymer concentration : $500$~ppm, $Q=3.75$~ml/min, $Pe=317$.) - Inset : a close-up at the scale of $4$ channels.}
\label{fig:fig5}
\end{figure}
 Some similar features are observed on the variation of the dispersion coefficient $D$ plotted 
 in figure~\ref{fig:fig5} as a function of the distance $X$ from the injection ($D$ is computed 
 from the value of $D/U^2$ given by the fit with $U$ equal to $X/{\overline t}$.) This time, 
 $D(X)$ is globally constant with $X$; like $\overline t$, it displays periodic oscillations
related to the structure of the network which will be discussed below. While these curves have been obtained for dyed
fluid displacing clear fluid, comparison experiments have been realized with clear fluid displacingg
dyed fluid; no systematic difference between the two sets of data was observed, confirming that
there are no buoyancy driven instabilities.\\
The oscillations of $\overline t$ and $D$ as $X$ varies are closely related to mixing at the 
junctions and to the exchange of tracers between the transverse channels and the rest of the
flow (a quantitative analyzis will be presented in section~\ref{secI}.)\\ 
The above analysis has been performed for all experiments realized with both polymer 
concentrations. In the following section,  variations with $Pe$ of the dispersion coefficient $D$ 
measured in this way  are discussed  
\subsection{Flow velocity dependence of dispersion coefficient}
\label{sec0}
\begin{figure}[h!]
\includegraphics[width=\W]{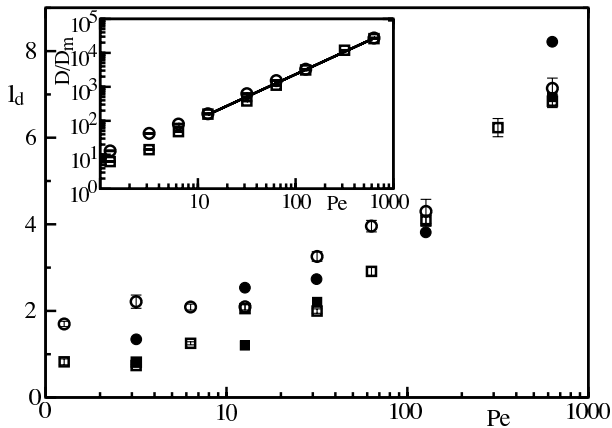}
\caption{Variation of the dispersivity $l_d$ (mm) with the P{\'e}clet number for experiments with 
water-polymer solutions : ($\square$),($\blacksquare$) : $500$~ppm concentration  - ($\circ$), 
($\bullet$) $1000$~ppm. Open (resp. grey) symbols : averaging interval : $35$ (resp. $0.4$) mesh 
sizes - Inset : variation of normalized dispersion coefficient  $D/D_m$ as function of the P{\'e}clet 
number for two polymer concentration (same symbols as in main graph). Solid line : power law fit 
for $Pe>10$ (exponent $1.35 \pm 0.03$).}
\label{fig:fig6}
\end{figure}
The variations with the P\'eclet number $Pe$ of the dispersion coefficient $D$ determined as explained in the previous section are displayed in the inset of figure~\ref{fig:fig6}.  For both polymer concentrations, 
 $D$ always  increases with  $Pe$ but two different variation regimes are visible. \\
For $Pe \ge 10$, values of $D$ corresponding to the two polymer concentrations
 fall on top of each other and 
 increase like $D \sim Pe^\beta$ with $\beta \simeq 1.35$. This is in good agreement with numerical 
 simulations~\cite{bruderer01} realized for a similar geometry and degree of heterogeneity (as 
 characterized by $\sigma(a)/\bar{a} \ = \ 0.33$) and for which a power law variation with an exponent of 
 the order of $1.3$ is also obtained.
 The value significantly higher than $1$ of this exponent cannot be accounted for solely by geometrical
dispersion (which would give a value of $1$) even if a logarithmic correction factor
 such that  $D~\propto~Pe~Log Pe$ is introduced~\cite{arcangelis86}. A likely 
 hypothesis is that this variation reflects a crossover from geometrical dispersion ($D \propto Pe$) to
 Taylor dispersion ($D \propto Pe^2$): in that range of P\'eclet numbers, both mechanisms would
 then contribute to the dispersion process. \\
For $Pe<10$, on the contrary,  values of $D$ obtained for the  $1000$~ppm polymer solution are 
higher than those obtained with the $500$~ppm one. The variation of $D$ 
with $Pe$ in that range has been studied more precisely by plotting   the dispersivity $l_d = D/U$ 
 in the main graph  of figure~\ref{fig:fig6}. The value of $l_d$ is approximately constant for
$Pe \ < \ 10$ for both polymer concentrations and is larger for the $1000$~pmm solution 
($l_d = 1.71$~mm) than for the $500$~ppm one ($l_d = 0.8$~mm). 
 This implies that, in this range of P\'eclet numbers, the geometrical mechanism 
controls dispersion and that the corresponding dispersivity increases with the shear thinning character
of the fluid (at higher $Pe$ values, $l_d$ is, in contrast, almost identical for the two solutions).\\ 
The values plotted in figure~\ref{fig:fig6} have been obtained by averaging the concentration $C$ 
over $35$ channels in the direction perpendicular to the mean flow. In order to 
 estimate the influence of the heterogeneities of the network, we performed an analysis in 
 which the concentration $C$  is only averaged over 
 $\Delta y = 4$ pixels (or about $0.55$ mesh sizes of the lattice): the coordinates $y$  of these measurement
lines are chosen so that all pixels are inside  longitudinal channels or  junctions and the dispersivity is determined as above. \\
These values of $l_d$ are represented as grey symbols in figure~\ref{fig:fig6} and are only 
slightly lower than those obtained for $\Delta y = 35$~mm. This shows that there are no
 large scale heterogeneities of the network, such as high or low permeability channels of width significantly larger than the mesh size which would increase the dispersivity.\\
Since all data points  correspond to $Pe \ge 1$, the {\it direct} influence on longitudinal dispersion of molecular  diffusion  is negligible: It has however a strong indirect 
influence at the lower P\'eclet numbers investigated ($Pe \lesssim 10$). The transit time of 
the tracer inside the junctions  or individual flow channels is then large enough so that molecular diffusion across the flow lines is significant : this influences strongly the redistribution of the
 incoming tracer  between channels leaving each junction~\cite{mourzenko02}. 
In the next section, we show that, in addition to the determination of macroscopic parameters like $U$, $D$ (or  $l_d$), the concentration maps  allow to investigate mixing  processes at the pore scale or even below.
\subsection{Tracer exchange dynamics between transverse and longitudinal channels.}
\label{secI}
\begin{figure}[h!]
\includegraphics[width=\W]{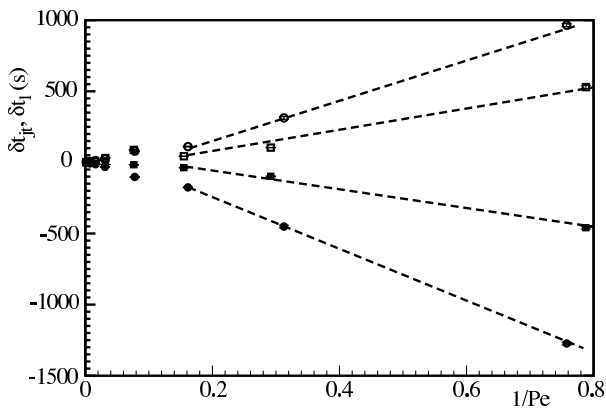}
\caption{Variation of $\delta t_{l}$ (filled symbols) and $\delta t_{jt}$ (open symbols)  with 
$1/Pe$. ($\square,\blacksquare$): experiments realized with $500$~ppm polymer solutions, 
($\circ,\bullet$): $1000$~ppm. Dashed lines are guides for the eyes.}
\label{fig:fig7}
\end{figure}

The variations with the distance $X$ from the inlet  of both the mean transit time
${\overline t}(X)$ (Fig.~\ref{fig:fig4}) and the dispersion coefficient $D(X)$ (Fig.~\ref{fig:fig5}) 
display periodic oscillations about respectively an increasing linear trend and a constant value.
In the following, the oscillations of $D$ will be characterized quantitatively by the difference 
between $D(X)$ and its mean value;  the variations of ${\overline t}(X)$ will be characterized by its difference with the linear regression ligne over all data points reflecting the mean front velocity $U$.
At a given distance $X$ the time corresponding to this regression is equal to $X/U$. 
The difference ${\overline t}(X)-X/U$ is negative when the line located at the distance $X$ from the inlet, and over which $C(X,t)$ is computed, contains only longitudinal channels parallel to the flow; it is positive when  the line contains  both transverse channels and junctions. Regarding $D(X)$, it  is also lower than the mean value when  the line  $X = cst.$ contains only longitudinal channels and higher  when it  contains both longitudinal and transverse channels.
The variations of ${\overline t}(X)$ reflect the different influence of longitudinal and
 transverse channels  on transport. As already discussed in section~\ref{dismech}, in a weakly disordered  square network like the present one, most convective flux is localized inside 
  the longitudinal channels  (parallel to the mean flow). 
The mean velocity inside them is then significantly higher than in the transverse channels and they
get saturated faster with the displacing fluid. There is therefore a time lag between the saturation
of the transverse and longitudinal  channels at a same distance $X$ from the inlet which explains
the oscillations of  ${\overline t}(X)$ in Figure~\ref{fig:fig4}.\\
Moreover, the respective amplitudes of the successive minimas and maximas of ${\overline t}(X)-X/U$
are found to be almost constant from one to the other. In the following, the time lags will therefore be characterized by the respecting averages $\delta t_{l}$ and $\delta t_{jt}$ over all 
minimal  and maximal values of ${\overline t}(X)-X/U$.\\
In the limit of a perfect mixing at the junctions ($Pe < 10$), fluid particles do not retain the memory of their past trajectory (ie whether they got previously trapped inside slow transverse channels). Therefore the time lag for $X\ =\ cst$ lines containing transverse channels should reflect directly the residence time in an individual (slow) transverse channel, weighted by the volume fraction corresponding to these channels. As long as $Pe \gg 1$ and molecular diffusion is negligible, the residence time  in a given channel will be inversely proportional to the local velocity; the latter is, in turn, proportional to the mean velocity $U$ as long as (as in the present case) the Reynolds number is low enough and the linear Stokes equation is approximately applicable. As a result, the local velocity in a channel is proportional to the mean velocity $U$ and the time lag  $\delta t_{jt} > 0$ should  vary as $1/U$ (for similar reasons 
$\delta t_l < 0$  should also vary as $1/U$).
Figure~\ref{fig:fig7} displays the variations of $\delta t_{jt}$ and $\delta t_l$ with $1/U$ for the two 
polymer solutions investigated : in both cases, the variation is indeed linear for $1/U \ge 40$~s.mm$^{-1}$ (corresponding to $Pe \le 6$). \\
For $1/U < 40$ ($Pe > 6$), $\delta t_{jt}$ (resp. $\delta t_l$) is higher  (resp. lower) than the values extrapolated from the linear  trend for $1/U \ge 40$ :  the transition is observed at the same mean velocity for the two solutions. 
This increase of the absolute values of the time lags 
reflects very likely the breakdown  of the assumption of perfect mixing at junctions or inside individual channels : at high  P\'eclet numbers, a solute particle may indeed flow through several junctions
and channels without moving across flow lines through transverse molecular diffusion. If the lattice is weakly disordered, the solute may then remain for a longer time inside a sequence of longitudinal high velocity channels  than in the case of a perfect mixing at the junctions : the corresponding value
of $\delta t_l$ will then be lower. Similarly a solute particle may remained trapped for a longer time.   in slow zones than if mixing was more effective at the junctions so that $\delta t_{jt}$ is increased.\\
Although the transition between the above two regimes takes place at the same P\'eclet number for both solutions, the absolute amplitude of the variations of $\delta t_{jt}$ and $\delta t_l$ with $1/U$ is significantly larger for the $1000$~ppm one.
This is direct consequence of the shear thinning properties of the fluid : 
 the effective viscosity of the solutions  increases much more with the polymer concentration in slow
transverse  channels (where the shear rate is low) than in fast longitudinal ones : as a result, the
contrast  between the velocities (and therefore the residence times) in the longitudinal and transverse channels is enhanced, leading to the observed increase of $\vert \delta t_{jt} \vert$ and $\vert \delta t_l \vert$. This enhancement
of velocity contrasts for shear thinning fluids is discussed in more detail in section~\ref{disc}.\\
At the opposite limit of low velocities such that $Pe < 1$ ($1/U \ge 250$~s$\times$mm$^{-1}$), 
 longitudinal diffusive transfer becomes significant.  The increase of the residence times with 
 $1/U$ is then limited by molecular diffusion to a value of  the order of a few  $l^2/D_m$ ($\sim 1000$s): the variations of  $\delta t_{jt}$ and $\delta t_l$ should then level off at high $1/U$ values. The lowest values of $Pe$ are however still too high in our experiments to observe this effect.\\
These results suggest therefore that the type of local observations reported here provide important 
informations on mixing processes at the pore scale. 
As pointed out recently~ \cite{grubert01,park01}, these processes may, in turn,  influence strongly
mass transfer at the macroscopic scale. In a similar perspective, we shall now investigate the dependence of the geometry of the iso concentration fronts on the flow velocity and the polymer concentration. We shall also discuss their relation to local mixing in the pores. 
\subsection{Geometry of tracer displacement fronts}
\label{sec:secII}
\begin{figure}[h!]
\includegraphics[width=\W]{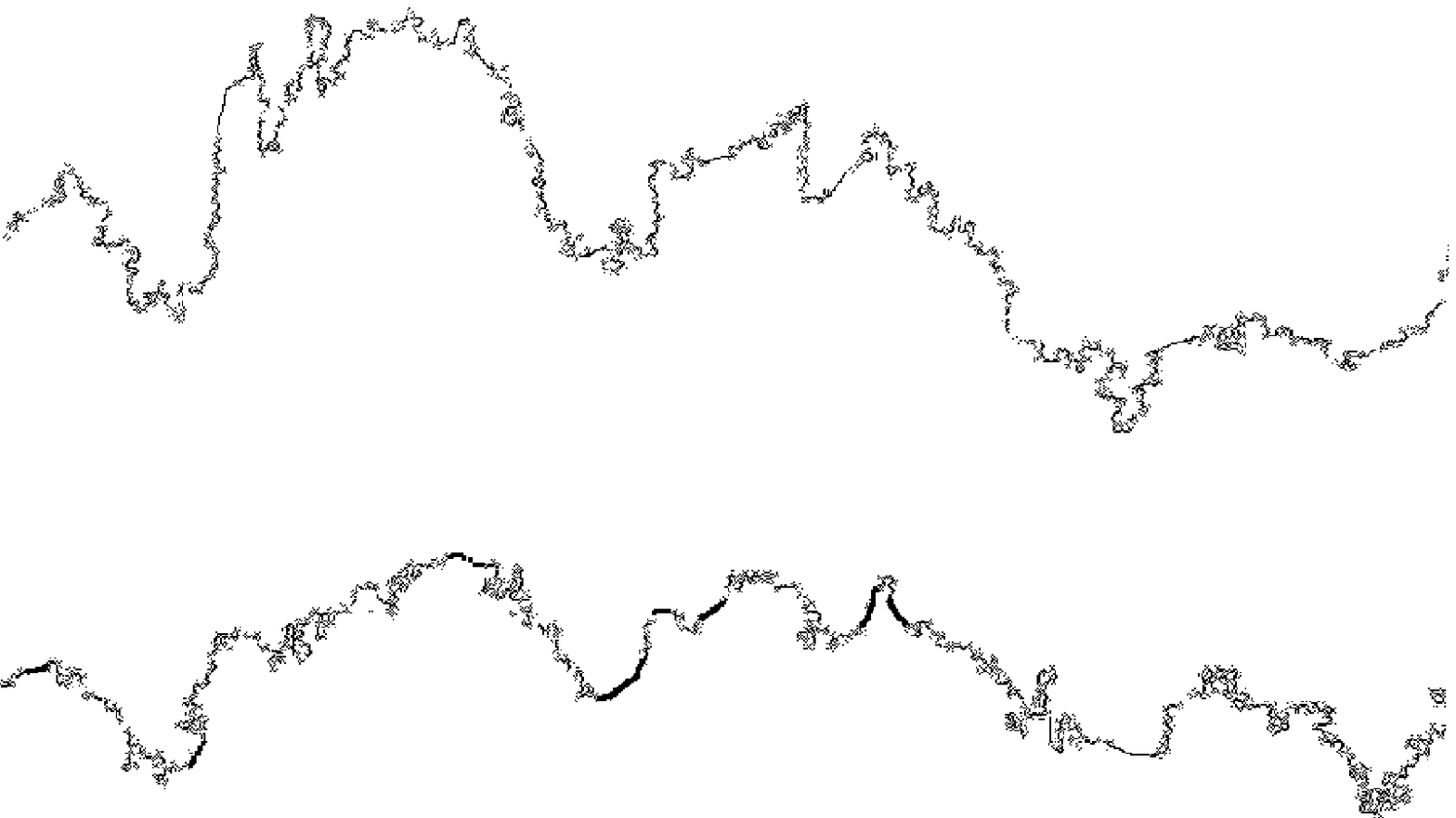}
\caption{Isoconcentration fronts at a mean velocity  $U=0.005$~mm/s ($Pe=1.3$) for polymer
 solutions of concentrations $1000$~ppm (upper curve) and $500$~ppm (lower curve).
  Mean distance of tracer from inlet : $0.5 \times  L$ ($L$ = model length). Flow is upward
   on the figure. Front widths :  $\sigma\ =\  4.5$~mm ($1000$~ppm) and $\sigma\ =\  2.6$~mm ($500$~ppm).}
\label{fig:fig8}
\end{figure}
\begin{figure}[h!]
\includegraphics[width=\W]{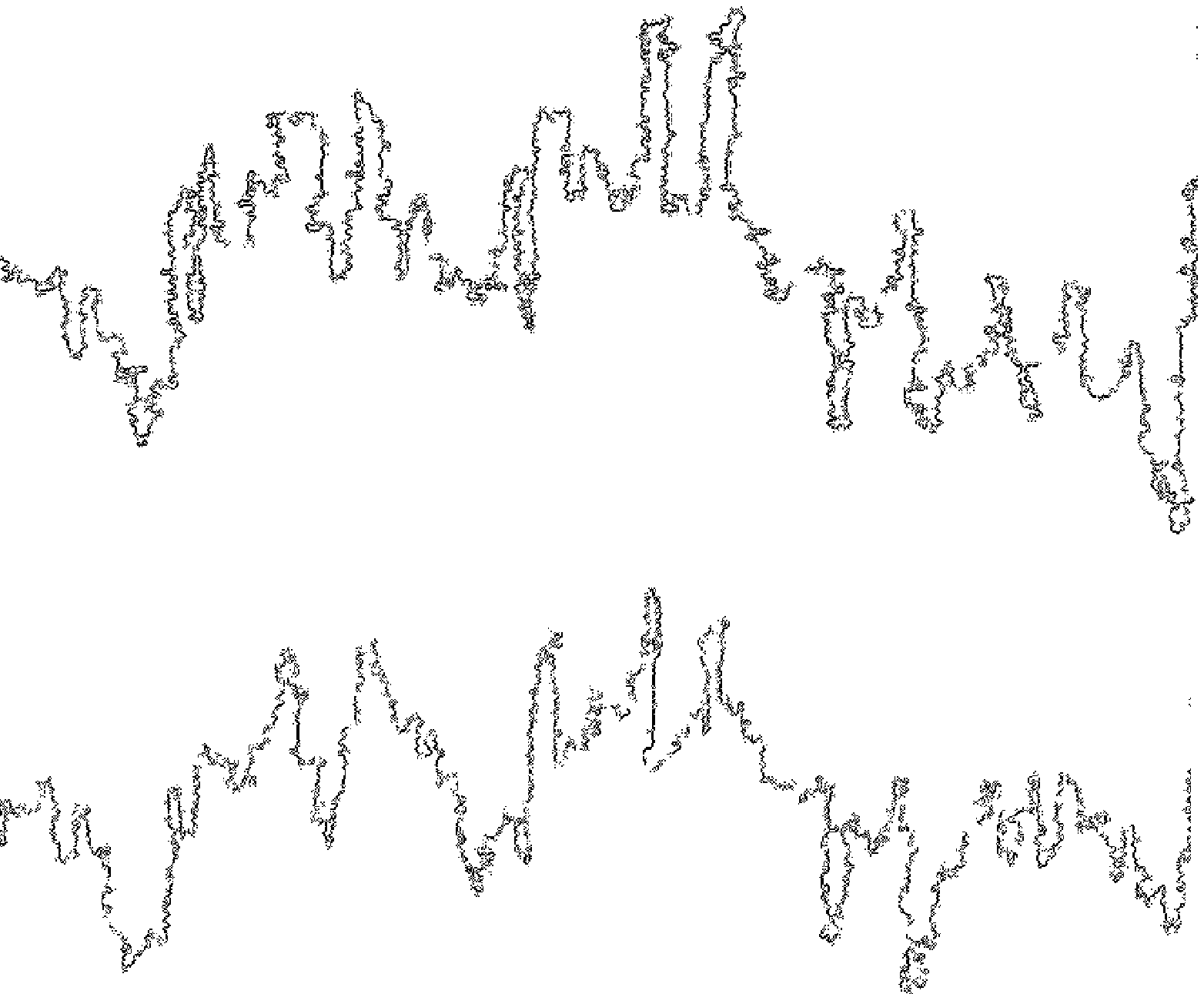}
\caption{Isoconcentration fronts for a mean velocity  $U=0.125$~mm/s ($Pe=32$) for polymer
 solutions of concentrations $1000$~ppm (upper curve) and $500$~ppm (lower curve). Mean
  distance of tracer from inlet : $0.5 \times  L$ ($L$ = model length). Flow is upward on the figure.
   Front widths  $\sigma\ =\ 6.1$~mm ($1000$~ppm) and $\sigma \ =\ 4.6$~mm ($500$~ppm).}
\label{fig:fig9}
\end{figure}
In the present  experiments, the pixel size in the concentration maps is   $0.4$ times the
 mean channel width. This allows for a study of the tracer distribution in the mixing 
 zone at length scales varying from the size of the network down to a fraction of the pore size. 
 For practical reasons, we shall not use the full spatial concentration distribution in the following 
 analysis : we chose instead to characterize its spatial heterogeneity by the isoconcentration lines 
 $c=0.5$, which are assumed  to reflect the  {\it displacement front} geometry: examples of such 
 fronts determined by a thresholding procedure are displayed in figures~\ref{fig:fig8} 
 and~\ref{fig:fig9} for P\'eclet numbers $Pe$ respectively equal to $1.3$ and $32$.\\
The width of the front  parallel to the mean flow 
is larger for the more concentrated solution and it increases with the P\'eclet number. Also, at 
high $Pe$ values, large spikes are visible while the front is relatively smooth at 
lower ones.  In spite of these differences, the main geometrical features of the front are similar: 
large peaks and troughs are generally located at the same points for different flow velocities and 
polymer concentrations. This confirms that irregularities of the front structure are associated to 
deterministic features of the velocity field and not to uncontrolled imperfections of the injection.\\
Quantitatively, the effective   width of the front  parallel to the flow is characterized in the following
 by the standard deviation $\sigma(\bar{x})$ of the distance $x$ of its points  from the inlet:
$\sigma(\bar{x})$ satisfies $\sigma(\bar{x})=<(\bar{x}-x(y))^2>^{1/2}$ in which $\bar x$ is the
mean value of $x$. In figures~\ref{fig:fig8} and~\ref{fig:fig9},
 $\bar x$ is equal to half the length of the model and the values of $\sigma$ corresponding to
 the curves displayed are listed in the captions.
\begin{figure}[h!]
\includegraphics[width=\W]{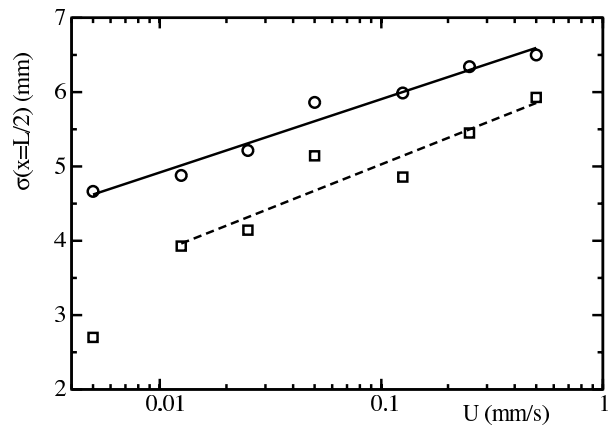}
\caption{Effective front width $\sigma$ as a function  of the mean front velocity $U$~(mm/s) 
for a mean distance  ${\bar x} = 0.5 \times L$ and polymer concentrations 
equal to $1000$~ppm ($\circ$) and $500$~ppm ($\square$). Solid (resp. dotted) lines : fit of 
the experimental data for a $1000$~ppm (resp. $500$~ppm) polymer solution by equation  
$\sigma = a+b \, log(U)$ with $a=6.9\pm0.1$ and $b=0.4\pm0.05$ (resp.  $a=6.2\pm0.2$ and 
$b=0.5\pm0.1$.}
\label{fig:fig10}
\end{figure}
Figure~\ref{fig:fig10} displays the variation of $\sigma$  as a function of the 
mean flow velocity $U$ for ${\bar x}\ =\ L/2$. For both solutions, the width $\sigma$  increases 
logarithmically with  $U$:  the value of $\sigma$ is  larger for the  $1000$~ppm solution
while   the slope of its variation with $U$ in figure~\ref{fig:fig10} is slightly lower.\\
The values of $\sigma$ are different for the two solutions  because the effective viscosity decreases faster with the velocity gradients for the $1000$~ppm solution than for the $500$~ppm one. 
The ratio between the effective viscosities, and therefore the velocities in the longitudinal 
and transverse channels is therefore higher and the front width which is directly related  to
this ratio increases.\\ 
The second major feature of the above results at high velocities is the large amplitude of the peaks and troughs observed on the front while they are smaller and narrower at lower velocities. This, too, may 
be explained by  the reduced tracer mixing at junctions  at high P\'eclet numbers (section~\ref{secI}) : solute remaining  for a long distance inside longitudinal, high velocity, channels will contribute to the spikes while that moving through a sequence of slow lateral channels will contribute to the troughs. 
At lower P\'eclet numbers, mixing is more efficient and solute particules sample more effectively the
velocity distribution: this reduces the dispersion of the transit times and, therefore the amplitude
of the peaks and troughs. 
\section{Discussion and conclusions}
\label{disc}
The local analysis of the transit times and of the front geometry have  therefore provided important informations 
on mixing inside junctions and flow channels and on its dependence on the P\'eclet number. 
This information greatly helps one to interprete  the macroscopic 
  dispersivity  measurements of section~\ref{sec:quant}.
Some of the features observed are specific to  2D systems while others 
can occur in usual 3D media.\\ 
At low P\'eclet numbers (typically  $Pe\ \le \ 10$), the dispersivity $l_d$ remains constant with $Pe$
and is lower for the $500$~ppm polymer solution than for the $1000$~ppm one. 
In  sections~\ref{secI} and~\ref{sec:secII}, we have seen that, in this regime,  transverse 
mixing in junctions and channels is very effective so that  the correlation length of the motion of solute particles is
of the order of the length $l$ of individual channels.
 As a result, this motion  
may be described as a sequence of random steps of varying durations and directions inside the 
medium; this is the geometrical dispersion regime discussed in section~\ref{dismech} 
and for which $D \propto Pe$ ($l_d\ = \ cst.$).\\
 In this regime, the factor of two difference  of the  dispersivities for  $500$ and $1000$~ppm
 solutions   is likely due to enhanced velocity contrasts  between 
the fast and slow flow regions. It is known, for instance~\cite{shah95},  that the mean flow 
velocity  inside a cylindrical channel under a given pressure gradient varies as the square of the 
radius $a$ for a Newtonian fluid and as  $a^{1+\frac{1}{n}}$ for a shear thinning fluid verifying 
equation~(\ref{powervisc}).
 Let us assume that the pressure gradient between the ends of flow channels  in parallel is constant. 
If  $\sigma(a) \ll \bar{a}$,  the standard deviation 
$\delta U$  of the mean velocities in the different channels should  scale like :
\begin{equation}
\frac{\delta U}{U} \sim  \frac{1+n}{n} \frac{\sigma(a)}{\bar{a}},
\label{eq:eq3}
\end{equation}
where $\sigma(a)/\bar{a}$ is the normalized standard deviation of the channel aperture 
(see section~\ref{subsec:model}). Still using  the same simplistic approach, the typical 
standard deviation  $\delta t$ of the transit time along a channel of length $l = 0.67$~mm 
should be $l/\delta U$. Estimating the dispersion coefficient $D$ from the relation
 $D\sim l^2/\delta t$ provides  the order of magnitude  of the dispersivity : 
\begin{equation}       
l_d\ \sim \ \frac{n+1}{n} \frac{\sigma_a}{a} \ l
\label{eq:eq4}
\end{equation}
Since $n$ decreases with the polymer concentration,  $l_d$ should therefore increase
for  a fixed aperture fluctuation  $\sigma(a)/a$.
Using in equation~(\ref{eq:eq4})  the values of $l$, $\sigma(a)/a$ and $n$ corresponding 
to the present experiments (table~\ref{tab:tab1})  leads to 
$l_d\simeq0.7$~mm and $1$~mm respectively for the $500$~ppm and $1000$~ppm solutions. 
These estimations 
are close to the experimental values $l_d \ = \ 0.8$~mm and $1.7$~mm  reported in 
section~\ref{sec0} for  the same solutions (figure~\ref{fig:fig6}).  The difference may be due
to the assumption of identical pressure gradients on different parallel channels  used to obtain 
equations~\ref{eq:eq3} and~\ref{eq:eq4}.\\
At higher P\'eclet numbers $Pe \ > \ 10$, $l_d$ is no longer constant but increases with $Pe$.
This reflects the  transition towards a second dispersion regime in which mixing is less effective. 
One must then take into account the stretching of dye parallel to the flow by
local velocity gradients in the flow section (dye moves slower in the vicinity of the walls than in
 the center of the channels). This stretching effect, is balanced by transverse molecular diffusion, 
resulting a Taylor-like dispersion mechanism (section~\ref{dismech}).
 This effect is made significant by the increase with  $Pe$ of the correlation length of dye transport
along chains of flow channels parallel to the mean flow discussed above in sections~\ref{secI} and~\ref{sec:secII}.\\
The effect of the local velocity gradients  is also enhanced by the specific topology of $2D$ micromodels. 
 The upper and lower walls are 
 indeed continuous and some flow lines  remain close to them over their full length: as in Taylor
dispersion, slow solute particles near these walls may only move away from them through 
molecular diffusion. Similarly, fast moving particles half way between the walls can only reach them
 through transverse molecular diffusion. The large correlation length of these velocity
contrasts also results in Taylor-like dispersion effects.\\
Yet, the influence of the disorder of the medium cannot be completely neglected (since some tracer always
moves into lower velocity transverse channels): the global dispersion results therefore 
from the combined effects of geometrical and Taylor dispersion.  As a result, the 
macroscopic dispersion coefficient $D$ follows in this regime  a power law $D \propto Pe^\beta$
of the P{\'e}clet number with an exponent  $\beta \simeq 1.3$ intermediate between the values $1$
and $2$ corresponding respectively to Taylor and geometrical dispersion.\\
Regarding the influence of the  shear thinning properties, increasing the polymer concentration enhances 
velocity contrasts between different flow channels while it flattens the velocity profile in individual channels.
The first effect increases geometrical dispersion and is indeed observed  at low P\'eclet numbers.
The second reduces Taylor dispersion : this explains why, at higher $Pe$ values, the values of $l_d$
 for the two polymer solutions are similar when  the influence of Taylor dispersion is large.\\
To conclude, the dispersion measurements reported in the present work for 
transparent micromodels  provide significant novel informations on the 
influence of the  flow velocity and fluid rheology on miscible displacements in porous media.
Quantitative high resolution optical  measurements have allowed for  thorough studies 
over a broad range of length scales: it has in particular been possible in the same experiment both to determine
macroscopic parameters such as the effective dispersivity and to analyse at the pore scale  the dynamics of 
concentration variations.\\
In particular, the local analysis of the  time lag between the invasions of longitudinal and transverse channels
of the model has allowed us to relate  the transition between two 
dispersion regimes for $Pe \simeq 10$ to variations of mixing in channel junctions. 
The variations of small scale structures of the displacement front with the P\'eclet number  and the
polymer concentration  also provides   informations on the spatial correlation of transport 
at the local scale. \\
In the future, investigation of these effects at still higher resolutions should allow for detailed direct
studies of mixing and flow patterns right inside individual flow channels and their 
junctions.

\begin{acknowledgments}
We thank C. Zarcone and the "Institut de M\'ecanique des
Fluides de Toulouse" for realizing and providing us with the
micromodel used in these experiments and G. Chauvin and R. Pidoux for realizing the 
experimental set-up. This work has been realized in the framework of the ECOS Sud 
program A03-E02 and of a CNRS-CONICET Franco-Argentinian "Programme International 
de Cooperation Scientifique" (PICS $n^o 2178$).
\end{acknowledgments}


\begin{thebibliography}{26}
\providecommand{\natexlab}[1]{#1}
\expandafter\ifx\csname urlstyle\endcsname\relax
  \providecommand{\doi}[1]{doi:\discretionary{}{}{}#1}\else
  \providecommand{\doi}{doi:\discretionary{}{}{}\begingroup
  \urlstyle{rm}\Url}\fi 
  
\bibitem{bear72}
{\normalsize J. Bear, ``Dynamics of Fluids in Porous Media", Elsevier
Publishing Co., New York (1972).}

\bibitem{dullien91}
{\normalsize F.A.L. Dullien, ``Porous Media, Fluid Transport and Pore
Structure", 2nd edition, Academic Press, New York (1991).}
 
\bibitem{charlaix88} {\normalsize E. Charlaix, J.P. Hulin, C. Leroy and C.
Zarcone. ``Experimental study of tracer dispersion in flow through two-dimensional networks of etched capillaries." {\it  J. Phys. D: Appl. Phys.} {\bf 21}, 1727 (1988).}

\bibitem{wong99}
{\normalsize P.Z. Wong, Ed. ``Methods in the physics of
porous media.", {\it Experimental methods in the
physical sciences} {\bf35}, Academic Press,  London (1999).}

\bibitem{zarcone83} {\normalsize R. Lenormand, C. Zarcone and A. Sarr.  ``Mechanism of the displacement of one fluid by another in a network of capillary ducts," {\it J.
Fluid Mech.} {\bf 135}, 337 (1983).}

\bibitem{Birovljev94}  {\normalsize A. Birovljev, K. J. M{\aa}l{\o}y, J. Feder, and
T. J{\o}ssang. ``Scaling structure of tracer dispersion fronts in porous media."
 {\it Phys. Rev. E} {\bf 49}, 5431 (1994).}

\bibitem{grubert01} 
{\normalsize D. Grubert. Effective dispersivities for a two-dimensional periodic fracture network 
by a continuous time random walk analysis of single-intersection simulations. 
{\it Water Resour. Res.}  {\bf 37}, 41 (2001).}

\bibitem{park01}
{\normalsize Y. Park, J.R. de Dreuzy, K. Lee and B. Berkowitz. ``Transport and intersection mixing
 in random fracture networks with power law length distributions." {\it Water Resour. Res.}
 {\bf 37}, 2493  (2001).}

\bibitem{sorbie89} {\normalsize
K.S. Sorbie, P.J. Clifford and E.R.W. Jones, ``The rheology of pseudoplastic fluids in 
porous media using network modeling." J. Colloid Interf. Sci. {\bf 130}, 508 (1989).}

\bibitem{paterson96} {\normalsize A. Paterson, A. D'Onofrio, C. Allain, J.P.
Hulin, M. Rosen and C. Gauthier. ``Tracer dispersion in a polymer solution flowing through a double porosity porous medium." {\it J. Phys. II France} {\bf 6}, 1639 (1996).}

\bibitem{freytes01}  {\normalsize M.A. Freytes, A. d'Onofrio, M. Rosen, C.
Allain, J.P. Hulin.  ``Gravity driven instabilities in miscible non Newtonian fluid displacements in porous media." {\it Physica A} {\bf290}, 286 (2001).}

\bibitem{mourzenko02}
{\normalsize V.V. Mourzenko, F. Yousefian, B. Kolbah, J.F. Thovert and P.M. and Adler,
``Solute transport at fracture intersections." {\it Water Resour. Res.}  {\bf 38}, 1000 (2002).}

\bibitem{berkowitz94}
{\normalsize B. Berkowitz, C. Naumann and L. Smith, ``Mass transfert at fracture intersections: 
An evaluation of mixing models."  {\it Water Resour. Res.}  {\bf 30}, 1765 (1994).}

\bibitem{taylor53} 
{\normalsize G.I. Taylor, ``Dispersion of soluble matter in solvent flowing slowly through a tube." 
{\it Proc. R. Soc. London A}  {\bf 219}, 186 (1953).}

\bibitem{aris56} 
{\normalsize R. Aris, ``On the dispersion of a solute in a fluid flowing through a tube." {\it Proc. 
R. Soc. London A}  {\bf 253}, 67 (1956).}

\bibitem{vartuli95} 
{\normalsize M. Vartuli, J.P. Hulin and G. Daccord, ``Taylor disper sion in a polymer solution 
flowing in a capillary tube." {\it AIChE J.} {\bf 41}, 1622 (1995).}

\bibitem{shah95}
{\normalsize C.B. Shah and Y.C. Yortsos, ``Aspects of flow of power-law fluids in porous media." 
{\it AIChE J.} {\bf 41}, 1099 (1995).}

\bibitem{fadili02}
{\normalsize A. Fadili, P. Tardyand and  A. Pearson, ``A 3D filtration law for power-law fluids 
in heterogeneous porous media."  
{\it J. Non-Newtonian Fluid Mech.} {\bf106}, 121 (2002).}

\bibitem{bruderer01}
{\normalsize C. Bruderer and Y. Bernab{\'e}, ``Network modeling of dispersion: Transition 
from Taylor dispersion in homogeneous networks to mechanical dispersion 
in very heterogeneous ones." {\it Water Resour. Res.}  {\bf 37}, 897 (2001).}

\bibitem{detwiler00}
{\normalsize R.L. Detwiler, H. Rajaram and R.J. Glass, ``Solute transport in variable-aperture fractures: 
An investigation of the relative importance of Taylor dispersion and macrodispersion."
{\it Water Resour. Res.} {\bf 36}, 1611 (2000).}

\bibitem{arcangelis86}
de Arcangelis L., J. Koplik, D. Redner and D. Wilkinson, Hydrodynamic dispersion 
in network of porous media, Phys. Rev. Lett., (1986) {\bf 57}, 996--999.

\end{thebibliography}

\end{document}